\documentclass[5p,times]{elsarticle}
\usepackage{graphicx}
\graphicspath{{images/}}
\usepackage{url}
\usepackage{amsmath}
\usepackage{listings}
\usepackage{comment}
\usepackage{color, soul}
\usepackage{microtype}
\usepackage{booktabs}

\soulregister\cite7
\soulregister\ref7
\soulregister\pageref7

\makeatletter
\def\ps@pprintTitle{%
  \let\@oddhead\@empty
  \let\@evenhead\@empty
  \let\@oddfoot\@empty
  \let\@evenfoot\@oddfoot}
\makeatother

\begin{document}

\begin{frontmatter}
\title{An AI-Based Solution for Secure Service Provisioning in IoT}
\author[address1]{Marco Arazzi}
\ead{marco.arazzi01@universitadipavia.it}
\affiliation[address1]{
organization={Department of Electrical, Computer and Biomedical Engineering, University of Pavia},
addressline={A. Ferrata, 5},
city={Pavia},
postcode={27100},
state={PV},
country={Italy}
}

\author[address1]{Mert Cihangiroglu\corref{correspondingauthor}}
\ead{mert.cihangiroglu01@universitadipavia.it}

\author[address1]{Serena Nicolazzo}
\cortext[correspondingauthor]{Corresponding author}
\ead{serena.nicolazzo@unipv.it}

\author[address1]{Antonino Nocera}

\ead{antonino.nocera@unipv.it}

\author[address3]{Vinod P.}
\ead{vinod.p@cusat.ac.in}
\affiliation[address3]{
organization={Department of Computer Applications},
addressline={Cochin University of Science and Technology},
state={Kerala},
country={India}
}

\date{June 2026}

\begin{abstract}
As the Internet of Things (IoT) continues its rapid expansion, the attack surface grows accordingly, with emerging threats constantly targeting smart objects and their interactions. In this evolving landscape, securing service provisioning is crucial to ensure the proper functioning, security, and reliability of the entire IoT ecosystem. Service provisioning encompasses key tasks such as device registration, configuration, authentication, authorization, and software deployment, all of which are essential for seamless and secure IoT operations. In this paper, we present a comprehensive framework designed to select the most suitable smart objects to deliver a target service within a given IoT environment while also monitoring the behavior of the entities involved during the service provisioning phase. To achieve this, we employ a Deep Reinforcement Learning (DRL) approach in which an intelligent agent learns, through interaction with a complex, dynamic environment, how to adapt to changes while adhering to predefined security constraints. For behavioral monitoring, we leverage Federated Learning (FL) to develop a global Behavioral Fingerprinting (BF) model that is fully distributed and can analyze how IoT devices interact within the network. In addition, the BF is used to compute a reliability score for each service provider, reflecting its degree of compliance with the defined security constraints. This score is then incorporated into the service provisioning process, allowing smart objects to select providers not only according to functional suitability but also to their reliability level. Finally, we conduct an extensive experimental evaluation to assess the robustness and scalability of our approach. The results demonstrate that our solution can be effectively deployed even on resource-constrained IoT devices, making it a viable and scalable security-enhancing mechanism for modern IoT ecosystems.
\end{abstract}


\end{frontmatter}

\section{Introduction}
The outbreak of the Internet of Things (IoT, hereafter) paradigm, characterized by a large number of heterogeneous smart devices and services, poses significant concerns about the security and privacy of all the actors of the system, including the exchanged data \cite{adat2018security}. Additionally, as we transition to Industry 5.0, the integration of IoT becomes even more critical in everyday life. Hence, securing IoT infrastructures, objects (i.e., devices in the IoT), and services is paramount to ensure safe human-machine collaboration \cite{arazzi2025privacy}.

In this context, the process of preparing, configuring and managing IoT devices, networks, and applications to deliver specific services to users or systems in a secure way is referred to as {\em secure service provisioning} \cite{alghamdi2019secure}. This includes several key tasks to ensure IoT devices can operate effectively and securely within a given ecosystem.
First, there should be a mechanism to securely add new IoT devices to the network and equipping them with unique identifiers.
Then techniques are required to provide the devices with the necessary parameters, such as network settings and security configurations, and to enable effective communication with other devices and systems. Furthermore, authentication and authorization strategies should be deployed to ensure that only legitimate entities can access the IoT network and its services with the proper roles and permissions.
In addition, the stages of software and services deployment and updates to fix any vulnerabilities are crucial.
Finally,  the performance and trustworthiness of IoT devices and services should be continuously monitored to address any issues that arise quickly.
Since issues such as unauthorized access, data breaches, malware attacks, and so on can alter the proper functioning and integrity of IoT devices, addressing various security threats during every step of service provisioning is a critical task and requires always-improved and robust security measures that do not compromise device performance.

To tackle these problems, we design a complete framework to {\em (i)} secure the selection of objects that serve as {\em service providers} in the IoT ecosystem according to several user-based security criteria and {\em (ii)} monitor the behavior of these objects throughout the service provisioning.
To solve the first challenge, we rely on the concept of Security Service Level Agreements (SecSLAs) that have been recently designed to standardize and assess the security and privacy requirements for smart objects interacting with Cloud applications \cite{rios2022security,casola2018security,nicolazzo2026service}.
Inspired by \cite{arazzi2024deep}, our work allows users to specify their security and privacy requirements before interacting with the IoT environment. These requirements are also considered to select suitable services from surrounding service providers.
Due to the complexity and high number of objects in the IoT ecosystem, in the first step of our approach, we rely on a heuristic strategy based on Deep Reinforcement Learning (DRL, hereafter). In particular, users are provided with an agent trained to take actions on their behalf and to choose, for every user request, the best service providers according to the security requirements expressed by the user. A tailored DRL reward function is employed to reward the agent based on the model parameters and the SecSLA exposed by the service provider.

Once the best provider is selected, service provisioning can start.
During this phase, our approach includes continuous monitoring to assess the trustworthiness of the service providers available in the IoT.
To do so, during the initial phase of our framework, we build the so-called {\em Behavioral Fingerprint} (BF) of devices that act as service providers using a set of useful features to identify
them \cite{aramini2022enhanced}. These features include both classical network identities (such as IP or MAC addresses) and the information from the packets that the device exchanges over the network. We aim to build a device profile modeling its typical behavior (how it interacts with the environment, patterns of communication, etc.) so that later we can measure the congruity of the current activities with the expected ones \cite{ferretti2021h2o,aramini2022enhanced}.
The BF computation proceeds through {\em(i)} the construction of behavioral models representing the expected conduct of every service provider in the network at the starting phase of our framework, and {\em (ii)} a subsequent monitoring activity to detect possible variations during the service provisioning.
Nevertheless, in existing distributed fingerprinting methods, an object constructs a model of a target contact's behavior using only data from its direct interactions. However, this provides a narrow perspective, as it ignores the services and messages exchanged between the target and other neighboring entities. However, creating a comprehensive model by aggregating information from multiple interactions across different nodes raises significant privacy risks. Hence, we borrow some ideas from a previous work \cite{arazzi2023fully}, which proposes computing the behavioral fingerprint of an object using a Federated Learning approach. Using this innovative ML strategy, our approach not only enhances the fingerprinting model with insights from multiple interactions across different nodes but also effectively addresses various security and privacy concerns related to the exchange of data among the involved entities.
In addition, the resulting BF model is also used to derive a reliability score for each service provider, capturing its degree of compliance with the expected behavior and security constraints. This score is then exploited as an additional input in the reward function of the DRL agent, together with user-defined security requirements, thus enabling the selection of service providers not only based on functional suitability, but also on their reliability.

Through an extensive experimental campaign, we demonstrate that our strategy is feasible, obtains a higher percentage of acquired services, and meets user security requirements. Moreover, we prove that our approach equips the nodes with the ability to detect if another peer is compromised before contacting it.
Interestingly, the proposed strategy is based on a lightweight behavioral fingerprint model suitable for modern powerful devices and more basic legacy devices.

The novel contributions of our paper can be summarized as follows.

\begin{itemize}
    \item We design a complete framework for secure service provisioning in IoT.
    \item We exploit the potential of SecSLA to empower users with the ability to decide their level of security and privacy needs, thus increasing their feeling of safety when it comes to smart devices and IoT technologies.
    \item We propose a framework for selecting the most suitable service provider by jointly considering security requirements and provider reliability, modeling the decision-making process as a Deep Reinforcement Learning (DRL) problem.
    \item We employ a lightweight behavioral fingerprinting model to monitor and assess the trustworthiness of IoT service providers during service provisioning.
\end{itemize}

The outline of this paper is as follows. In Section \ref{sec:related}, we describe the literature related to our approach. Section \ref{sec:background} is devoted to describing some basic concepts related to Security Service Level Agreements and Behavioral Fingerprints, which are useful in comprehending our proposal. In Section \ref{sec:description}, we overview our reference IoT model and examine the proposed framework. Section \ref{sec:experiments} deals with the presentation of the set of experiments carried out to test our approach and show its performance. Finally, Section \ref{sec:conclusion} draws our conclusions and presents possible future works related to our paper.

\section{Related Work}
\label{sec:related}
With the increase in the number of IoT devices, the demand for secure and high-quality service provisioning becomes a critical issue driving the attention of the scientific community. In this section, we review research related to the provision of services in the IoT.
In \cite{deng2018composition} the authors propose an IoT
service provision system utilizing a Mobile Edge Computing (MEC) model. In this model, multiple edge servers are deployed alongside access points through wireless networks. Using cached services on these edge servers can minimize latency and offload computational tasks, enhancing overall efficiency. The work \cite{niemirepo2015service} focuses on enabling automated provisioning of IoT services through a Service Oriented Device Architecture (SODA) to authorize new service providers who can
dynamically publish their innovative services on the system.
Zhao et al. \cite{zhao2016event} present an event-driven service provisioning mechanism for the interaction of the IoT system utilizing a multilevel and multidimensional model-based service
provisioning platform. This system enables
 access to large-scale heterogeneous resources and exposes
resource capabilities as light-weighted service interfaces.

All of the approaches mentioned above focus on various characteristics of IoT service provisioning, proposing frameworks that neglect the security aspect in their proposal.

Few works deal with secure service provisioning in the IoT \cite{khan2017towards,alghamdi2019secure,shahidinejad2023blockchain,shahidinejad2023blockchain,kazim2018framework}.
In \cite{khan2017towards}, the authors propose SSServProv, a framework for end-to-end secure and privacy-aware service provisioning within smart cities. Their approach involves registering service providers with a governmental domain authority to ensure compliance and participation in the system. The framework incorporates authentication, authorization, and a lightweight secure communication protocol to facilitate both data acquisition and service provisioning securely. To validate the security of these protocols, they utilized Scyther\footnote{\url{https://people.cispa.io/cas.cremers/scyther}}, an automated security verification tool.
Similarly, the systems designed in \cite{alghamdi2019secure,shahidinejad2023blockchain}
are two secure service provisioning schemes based on Blockchain. In particular, Alghamdi et al. \cite{alghamdi2019secure} proposes a framework that takes advantage of {\em (i)} a fair payment system for IoT Lightweight
Clients based on Blockchain and {\em(ii)} an incentive mechanism based on reputation. The study in \cite{shahidinejad2023blockchain} introduces a self-certified key exchange protocol leveraging Blockchain technology, specifically designed for hybrid electric vehicles operating within a fog computing environment. Kazim et al. \cite{kazim2018framework} propose a framework for dynamic and secure access to IoT services across multi-cloud environments, using an on-demand cloud model for authentication and Service Level Agreement (SLA) management. In addition, their system includes a service matching mechanism that identifies the most suitable IoT service based on user requirements on multiple external clouds. However, the framework does not incorporate security or privacy requirements in the matchmaking process.

Our proposal is different from the above ones, indeed, in our case the user gives some security requirements and the agents choose the best object in the environment providing them, hence, our aim is not to design a middleware to provide additional security functionalities. Our framework does not include a governmental domain authority, and the monitoring aspect of the service provider is performed also during the service provisioning through the mechanism of Behavioral Fingerprinting and is fully distributed.

\section{Background}
\label{sec:background}
In this section, we present the main concepts that can be useful for a clear understanding of our approach. In particular, we focus on the definitions of Security Service Level Agreement (SecSLA), Federated Learning (FL), and Behavioral Fingerprint (BF).

Table \ref{tab:SystemSymbols} summarizes the acronyms used in this paper.

\begin{table}
\centering
  \caption{Summary of the acronyms used in the paper}
  \begin{tabular}{ll}
\hline
    \textbf{Symbol} & \textbf{Description}\\
\hline
    BF & Behavioral Fingerprint\\
    DRL & Deep Reinforcement Learning\\
    FL & Federated Learning\\
    IoT & Internet of Things\\
    ML & Machine Learning\\
    SecSLA & Security Service Level Agremeent\\
    SLA & Service Level Agremeent\\
\hline
\end{tabular}
\label{tab:SystemSymbols}
\end{table}

\subsection{Security Service Level Agreements}

In present IoT environments, guaranteeing security and transparency remains a major challenge due to {\em(i)} the increasing number of interconnected devices offering diverse services, {\em(ii)} the complexity of new and heterogeneous architectures, and {\em(iii)} the increasing risk of data breaches and service disruptions. Although security has long been recognized as a potential attribute of Service Level Agreements (SLAs)~\cite{henning1999security}, traditional SLAs primarily focus on quantitative and measurable performance indicators (such as service availability, response time, and Quality of Service), not explicitly incorporating security metrics, leaving a gap in ensuring security commitments. Only recently, Security SLAs (SecSLAs) have been introduced as a step toward embedding security considerations into service provisioning, enabling a more structured and transparent approach to managing security in IoT infrastructures \cite{nicolazzo2026service}.

A Security SLA (SecSLA) has been defined by a report from the European Union Agency for Network and Information Security (ENISA) \cite{enisa2009enisa} as a contract between service providers and service customers stating the level of security granted between the parties \cite{casola2016automatically}. According to Chan et al., \cite{chan2004role}, the following security attributes, along with their corresponding metrics, play a crucial role in the definition and enforcement of SecSLAs, ensuring robust protection and accountability in service agreements.

\begin{itemize}
    \item \textbf{Availability} refers to maintaining uninterrupted access to network services, ensuring protection against Denial-of-Service (DoS) attacks. It is typically measured by the percentage of downtime caused by security incidents.
    \item \textbf{Data confidentiality} ensures that sensitive data remains protected from unauthorized access or disclosure.
    \item \textbf{Data integrity} protects data against unauthorized modifications or tampering attacks. A possible metric for this property is the percentage of integrity breaches detected.
    \item \textbf{Access control} restricts system access to authorized users and devices, ensuring that only permitted entities can interact with network elements, services, and applications.
    \item \textbf{Authentication} is essential to verify the identity of communicating entities and prevent spoofing attacks.
    \item \textbf{Non-repudiation} guarantees that actions, messages, or transactions can be traced back to their origin, providing accountability. A common metric for this property is the percentage of transactions secured using digital signatures.
    \item \textbf{Communication security} ensures that data is transmitted exclusively between authorized endpoints. One way to quantify this property is to measure the percentage of hijacked session time.
    \item \textbf{Privacy} focuses on protecting sensitive information and identities from unauthorized exposure or misuse.
\end{itemize}

\begin{figure}[ht]
    \centering
    \includegraphics[scale=0.45]{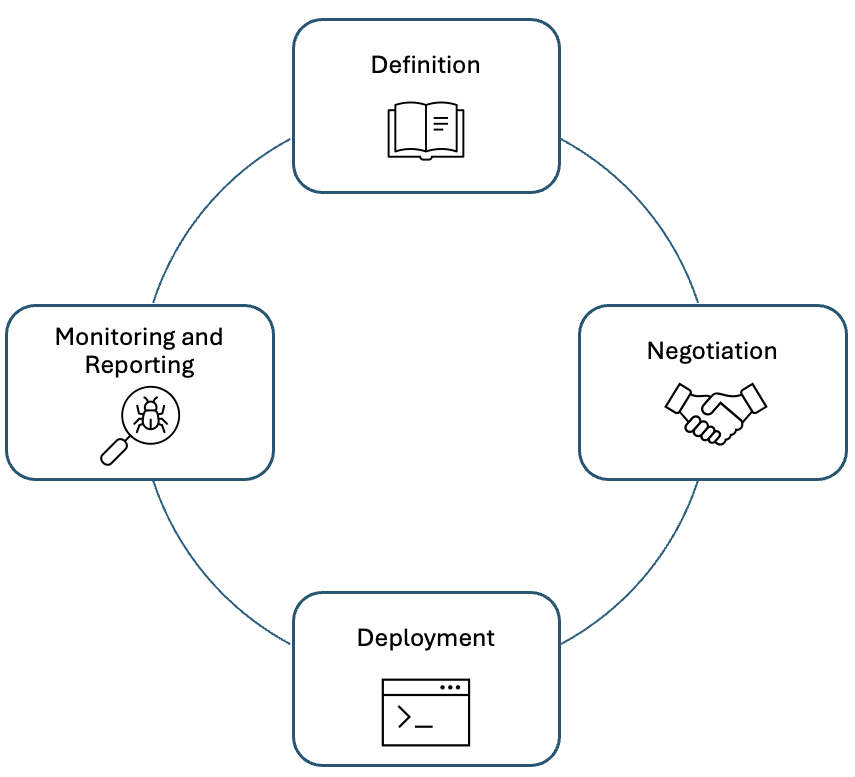}
    \caption{SecSLA life cycle}
    \label{fig:SecSLALifecycle}
\end{figure}

Similar to traditional SLAs, the Security SLA (SecSLA) lifecycle, shown in Figure \ref{fig:SecSLALifecycle}, consists of the following key stages, each defining a specific responsibility for the customer or the service provider.

\begin{enumerate}
    \item {\bf Definition}. This initial phase involves specifying the security parameters and metrics that will be included in the SecSLA.
    \item {\bf Negotiation}. In this stage, the various stakeholders in the system establish security requirements.
    \item {\bf Deployment}. During this phase, the necessary security services are implemented through the deployment of appropriate security mechanisms.
    \item {\bf Monitoring and Reporting}. Once the SecSLA is in effect, continuous monitoring is performed to ensure compliance. During this last phase, multiple actions can take place, namely {\em (i)} reporting of security and performance levels, {\em (ii)} prediction of contract violations, {\em (iii)} management of corrective actions to maintain security compliance, and {\em (iv)} implementation of incident response and remediation plans.
\end{enumerate}

The SLA and SecSLA contracts have been mostly written
using natural expressions, and compliance examination was carried out manually \cite{keller2003wsla}. The current most prominent industrial approaches for SLA language specification are:
WSLA, WS-Agreement, SLA$*$, CSLA, SLAC, RBSLA, and SLA-IoT \cite{maarouf2015review}. In our approach, we refer to the WSLA framework proposed by IBM \cite{ludwig2003web,bianco2008service}. Primarily, the WSLA allows the creation of machine-readable SLAs for services implemented using Web services technology, its language is based on XML and defined as an XML schema.

\subsection{Federated Learning}

Federated Learning (FL) is a machine learning approach that enables model training in a distributed fashion across multiple devices, each maintaining its private local dataset.

In this framework, key participants include $\mathcal{C}$ devices, referred to as ``clients'', which perform local training while keeping their data private. Additionally, there is a central server, known as the ``aggregator'', responsible for coordinating the FL process by collecting and aggregating local model updates from the clients.
Specifically, FL trains a global model $\mathbf{w}$ by uploading the weights of local models
$\{\mathbf{w}^i|i \in \mathcal{C}\}$ to a parametric server optimizing a loss function:

$$\min\limits_{\mathbf{w}} l(\mathbf{w}) = \sum_{i=1}^n{\frac{s_i}{\mathcal{C}}L_i(\mathbf{w}^i)}$$

\noindent
where $L_i(\mathbf{w^i})= \frac{1}{s_i}\sum_{j \in I_i}{l_j(\mathbf{w}^i, x_i)}$ is the loss function, $s_i$ is the local data size of the {\em i}-th client, and $I_i$ identifies the set of data indices
with $|I_i|=s_i$, and $x_j$ is a data point.

\begin{figure}[ht]
    \centering
    \includegraphics[scale=0.6]{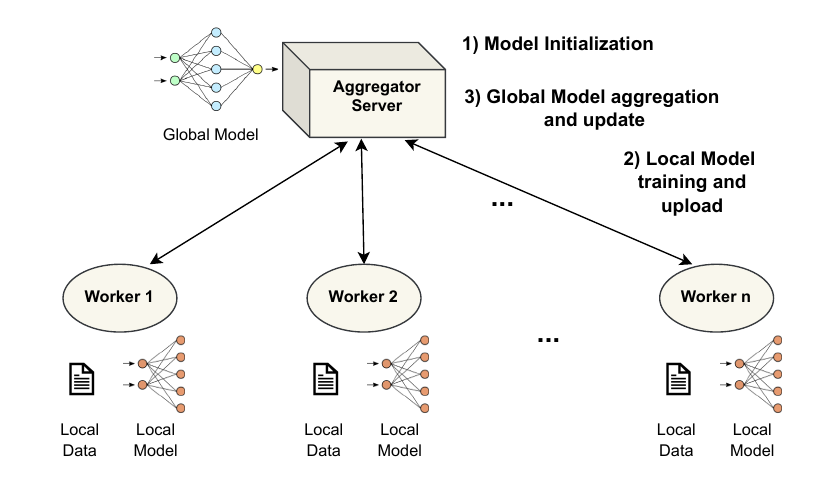}
    \caption{Federated Learning basic workflow}
    \label{fig:FL}
\end{figure}

The basic FL workflow \cite{zhang2021survey}, shown in Figure \ref{fig:FL}, follows these steps:

\begin{enumerate}
    \item Model initialization, in which the central server initializes all the necessary parameters for the global ML model $\mathbf{w}$. This phase also includes the client selection process.
    \item Training and upload of the local model, in which the clients download the current global model and perform local training using their private data. After that, each client computes the update of the model parameters and transmits them to the central server. Local training typically involves multiple iterations of gradient descent, back-propagation, or other optimization methods to improve the local model's performance. Specifically, at the {\em t}-iteration, each client updates the global model by training with their datasets: $\mathbf{w}^i_t \leftarrow \mathbf{w}^i_t - \eta \frac{\partial L(\mathbf{w}_t,b)}{\partial \mathbf{w}^i_t}$, where $\eta$ and $b$ identify the learning rate and local batch, respectively.
    \item Global model aggregation and update, in which the central server collects and aggregates the updates of the model parameters from all clients $\{\mathbf{w}^i|i \in \mathcal{C}\}$. The central server can employ various aggregation methods like averaging, weighted averaging, or secure multi-party computation to incorporate the received updates from each client, thus improving the performance of the global model.
\end{enumerate}

\subsection{Behavioral Fingerprints}
With the term Behavioral Fingerprint (BF) of an IoT object, we refer to a unique and identifiable pattern of the object's normal activity and interactions within a network. This metric is derived from analyzing more application-level features that model objects' traits. Among these characteristics, there are: protocols, request-response sequences, communication patterns, resource usage, and response to specific triggers and any periodicity in specific typologies of packets along with their sizes \cite{ferretti2021h2o,aramini2022enhanced}. In particular, the approach we follow \cite{aramini2022enhanced} for the computation of BF considers the following parameters.

\begin{itemize}
    \item \textbf{Source Port Type} refers to the category of the source port (user, system, or dynamic) from which an IoT device transmits a packet. Typically, IoT devices use predefined ports designated by the manufacturer for communication.
    \item \textbf{TCP Flag} is the flag embedded in each packet that indicates its purpose, such as SYN, SYN-ACK, PUSH, or FIN.
    \item \textbf{Encapsulated Protocol Types} specifies the communication protocol used by an IoT device for packet transmission, such as TCP, UDP, ICMP, or IGMP.
    \item \textbf{Inter-Arrival Time (IAT) of Consecutive Packets} measures the time interval between two successive packets sent by an IoT device. Typically, IoT devices transmit packets at a predefined and consistent frequency.
    \item \textbf{Packet Length} represents the size of a packet in bytes. Generally, IoT devices send structured messages with relatively constant sizes.
    \item \textbf{Payload Value} denotes the actual data measured by IoT devices and carried within transmitted packets. Incorporating this feature allows the system to assess the validity of reported measurements, as anomalies may cause IoT sensors to transmit incorrect or abnormal values.
    \item \textbf{Payload Value Shift} represents the difference between the payload value of the current packet and that of the previous packet. Tracking this variation helps detect anomalies, as certain attacks or malfunctions may cause abrupt or continuous fluctuations in sensor readings.
\end{itemize}

In practice, given a node $d_y$ in a network $IoT$, the Behavioral Fingerprint $BF_{d_y}$ is calculated as:

$$BF_{d_y}= \{ f_{d_x,d_y}|d_x,d_y \in IoT \land \langle d_x, d_y  \rangle \in C \}$$

\noindent
where:
\begin{itemize}
    \item $f_{d_x,d_y}$ is the fingerprint model that $d_x$ associates with $d_y$;
    \item $IoT = \{ d_i = i = 1, \dots, N \}$ is the reference IoT network of $N$ devices;
    \item $\langle\ d_x, d_y \rangle = \{pkt_1, pkt_2, \cdots, pkt_n\}$ is a sequence of packets;
    \item $C = \{ \langle\ d_x, d_y \rangle| d_x, D_y \in IoT\}$ is the set of all communications of a network $IoT$.
\end{itemize}

The approach maps the sequence of packets $\langle\ d_x, d_y \rangle$ to a sequence of symbols $\{s_1, s_2, \dots, s_n\}$. The fingerprint model $f_{d_x,d_y}$ consists of a neural network model that is trained to predict a symbol in the sequence based on the $m$ preceding symbols. Using BF, organizations can enhance security, privacy, and operational efficiency in IoT environments, ensuring that devices act as expected and mitigating risks from compromised or rogue IoT objects. Indeed, $BF_{d_y}$ can be used to assess whether the behavior of $d_y$ is unchanged.

\section{Description of our Approach}
\label{sec:description}
In this section, we provide a detailed description of our framework. As stated in the Introduction, our approach to IoT secure service provisioning consists of the following two main components.

\begin{itemize}
    \item \textbf{Secure Service Selection}. During this phase, the most suitable smart objects are selected to deliver a target service within a given IoT environment. A DRL approach is exploited, thanks to which an intelligent agent (deployed in an object that acts as a client) continuously interacts with a dynamic and complex environment, learning from its experiences and adapting to changes while adhering to predefined security constraints imposed by the user.
    \item \textbf{Secure Service Monitoring}. The service providers involved in service provisioning are continuously monitored through their Behavioral Fingerprints. These BFs are computed in the starting phase of the system via a Federated Learning approach.
\end{itemize}

The actors of our systems are the following.

\begin{itemize}
    \item \textbf{IoT nodes}, these nodes are client nodes that interact with different service providers to satisfy their user needs. They can select the best service according to some user protection requirements leveraging a DRL strategy.
    \item \textbf{Service providers} are IoT nodes capable of offering specific services. During the initial phase of our framework, these nodes actively participate in the Federated Learning process to compute their Behavioral Fingerprint.
    \item \textbf{Aggregator}, this node is in charge of aggregating the local contribution of the clients to compute a Global Model for the Behavioral Fingerprint computation.
\end{itemize}

In the following sections, we will describe these two steps of our framework in detail.

\subsection{Service Selection}
\label{sub:ServiceProvisioning}

\begin{figure}[ht]
    \centering
    \includegraphics[scale=0.6]{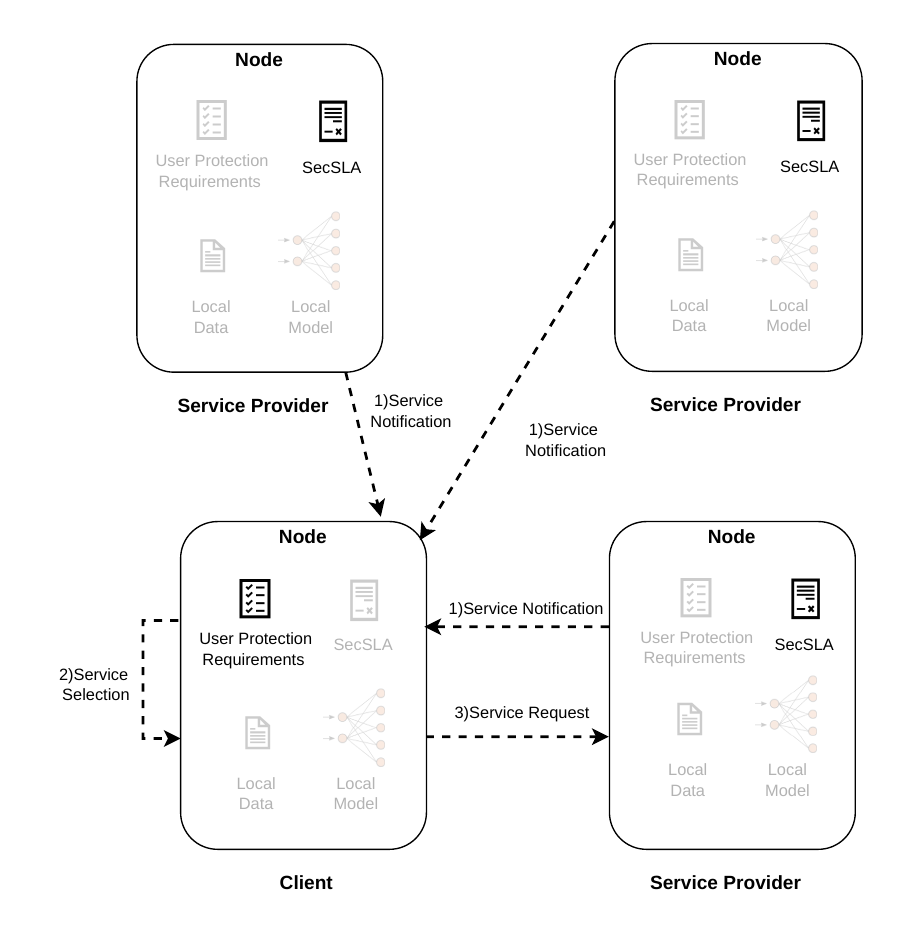}
    \caption{Service Provisioning workflow}
    \label{fig:ServiceSelection}
\end{figure}

During the {\em Service Selection} step of our framework, the client explores the environment to find suitable services in order to satisfy her/his user requests.
As shown in Figure \ref{fig:ServiceSelection}, the client node is equipped with some protection requirements expressed by the user at the initial phase of our framework. The service provider instead owns a machine-readable SecSLA (whose main characteristics have been described in Section \ref{sec:background}). We borrow some ideas from the work proposed by Arazzi et al. \cite{arazzi2024deep}, we can represent both the User Protection Requirement and the SecSLA in terms of {\em security class}, which is a set composed of several abstract security properties along with the corresponding security labels.
In particular, abstract security properties are directly derived from the CIA triad (Confidentiality, Integrity, and Availability). Each property can be associated with one of the following three security labels that correspond to the High (HC,
HI, and HA), Medium (MC, MI, and MA), and Low (LC, LI,
and LA) values. Again, such labels encode the strength of the user requirement or the extent to which the provided service is compliant with the reference property, respectively. Hence, $\mathcal{L}^C=\{HC, MC, LC, -\}$, with $HC \succ^C MC\succ^C LC\succ^C -$, is the set of labels associated with property $C$;
$\mathcal{L}^I=\{HI, MI, LI, -\}$, with $HC \succ^I MC\succ^I LC\succ^I -$, is the set of labels associated with the property $I$; finally,
$\mathcal{L}^A=\{HA, MA, LA, -\}$, with $HC \succ^A MC\succ^A LC\succ^A -$, is the set of labels associated with the property $A $.
An example of a {\em security class} $c_i$ that focuses on the properties of the CIA triad can be $c_i=[HC, HI, HA]$.
Furthermore, a security class dominates another one if and only if the dominance relationship holds for each of its components. For example, $[HC, HI, HA]\succeq[HC, MI, HA]$.
In addition, the distance between two security classes is computed as the sum of the distances between the corresponding components.

Thanks to this formalization, for a given service $s$, we can define the Euclidean distance $\delta^s$ between the two vectors referring to the User Protection Requirement $c^s_{UPR}$ and the SecSLA $c^s_{SecSLA}$, which can be computed and used as a parameter to quantify the similarity between a service and a user requirement.

$$\delta^s(c_{SecSLA},c_{UPR}) = \sqrt{\sum_{i=1}^m(c^s_{SecSLA}[A_i]-c^s_{UPR}[A_i])^2}$$

\noindent
where $c^s_{SecSLA}[A_i]$ (and $c^s_{UPR}[A_i]$) is the security label associated with a property related to a SecSLA (and to a User Protection Requirement) for a given service $s$, respectively. This metric will be used during the phase of service selection of the DRL algorithm by the client node to compute and select the best service among the encountered ones through a DRL algorithm.

In particular, Deep Reinforcement Learning (DRL) is an advanced Machine Learning paradigm that combines Reinforcement Learning (RL) with Deep Learning (DL) to allow an agent to learn optimal behaviors in a complex environment through interaction and trial-and-error \cite{arulkumaran2017deep}. In our approach, as the client object interacts with the environment, it encounters providers offering services relevant to the user. When this occurs, the client must decide whether to acquire the service based on several factors, namely, whether it is needed by the user and the security requirements expressed by the user.

At this point, the client has the following two possible actions to choose from.

\begin{enumerate}
    \item Acquire the service offered by the provider it has just encountered.
    \item Take no action, which means that it does not interact with  the current provider.
\end{enumerate}

In DRL, the reward function $\mathcal{R}(t)$ defines the criteria for assigning rewards based on the agent's state and the actions chosen and provides feedback to the agent about the effectiveness of its choices to achieve a given goal. In our framework, we define the reward associated with the acquisition of a service $s$ as a combination of compliance with user-defined security requirements and trustworthiness of the service provider derived from a Behavioral Fingerprinting (BF) model.

$$
\mathcal{R}^s(t) = \alpha \left(1 - \frac{\delta^s(c^s_{SecSLA},c^s_{UPR})}{\Delta} \right) + (1-\alpha)\cdot T_s
$$

\noindent
where $\delta^s(c_{SecSLA},c_{UPR})$ is the Euclidean distance between the vectors representing the User Protection Requirement $c_{UPR}$ and SecSLA $c_{SecSLA}$ for a given service $s$, and $\Delta$ is the maximum possible distance between two security classes. The term $T_s \in [0,1]$ represents the trustworthiness score of the service provider, calculated through the Behavioral Fingerprinting model and reflecting its degree of compliance with expected behavior and security constraints (see Section \ref{sub:ServiceMonitoring} for details). The parameter $\alpha \in [0,1]$ controls the trade-off between the security requirement matching and the reliability of the provider.

This formulation ensures that services that both satisfy user security requirements and exhibit high trustworthiness yield higher rewards, while services that poorly match user requirements or show unreliable behavior are penalized.

As the agent interacts with the environment, the objective is to learn the distribution of available services and associated operations to estimate the likelihood of encountering better service offers in future steps. Consequently, at any given step, the agent can reject the current service offer.

When the agent decides not to acquire a service, a small constant negative reward $\epsilon \leq 0$ is assigned. Since the negative price to pay for this action is negligible, the agent is encouraged to explore, but eventually, the accumulation of negative values pressures it to accept a provisioning.

$$
\mathcal{R}_{\text{no-action}}(t) = \epsilon
$$

Figure \ref{fig:ServiceSelection} shows the different actors and interactions during the {\em Service Selection} phase of our approach.

\subsection{Service Monitoring}
\label{sub:ServiceMonitoring}

This section describes the Federated Learning (FL) strategy used for computing Behavioral Fingerprinting models for the monitoring phase of our approach.

As stated in Section \ref{sec:background}, FL is a distributed and collaborative ML approach that enables model training across multiple decentralized devices while keeping local data private, eliminating the need to share raw datasets \cite{konevcny2015federated}. In recent years, this paradigm has gained attention for its potential to develop intelligent and privacy-preserving IoT applications \cite{nguyen2021federated,sanchez2021survey}.
Thanks to FL we can construct a comprehensive BF for a service provider that takes into account several interactions over the network, even across different services, combining the perspectives of different objects. Our strategy for the service monitoring phase consists of two main steps.

The first is {\em BF training phase} in which the models related to different service providers are trained. During this step, for each service provider to be monitored, a set of {\em workers} (with whom the service provider has to interact) should be selected. Observe that the initial phase in which behavioral fingerprinting models are built should be considered safe, i.e., no node is attacked or compromised.
This assumption reflects common commissioning practice in industrial automation, where the physical configuration and control software are tested and validated before final integration and production deployment. Our analysis targets precisely this phase, during which the network is active, but still managed and controlled by the system administration.
Then, to compute the BF we adopt the approach outlined in \citep{arazzi2023fully}, integrating payload-based features into our strategy. These additional features play a crucial role in defending against cyber-physical attacks, where adversaries attempt to manipulate sensor data to disrupt the normal operation of cyber-physical systems. Beyond payload-based features, this approach also incorporates traditional network parameters to characterize an object's behavior. These include source port type, TCP flags, encapsulated protocol types, inter-arrival time between consecutive packets, and packet length, alongside features derived from the payload. The method then maps the sequence of exchanged packets into a symbolic representation and applies a Gated Recurrent Unit (GRU) neural network for processing. The GRU model consists of two layers with 512 and 256 neurons, followed by a fully connected layer of 128 neurons and an output classification layer. A GRU-based architecture was selected over more complex models (such as LSTMs) to strike a balance between classification accuracy and computational efficiency, ensuring that behavioral fingerprinting models for IoT nodes remain both effective and scalable. The primary goal of the deep learning model is to predict the next symbol in a sequence, given a defined input window of past symbols. The approach continues with the aggregator that initializes a Global Model with randomly assigned learning parameters. Next, each client node connects to the aggregator to retrieve the current model. Once received, each worker updates its local model using its own dataset, which has been collected through direct interactions with the target node. During each training epoch, after computing its Local Model update, the worker transmits its contribution to the aggregator, which is responsible for aggregating all local updates. This process continuously refines the Global Model, improving its performance while preserving data privacy and preventing information leakage. These last two steps are repeated iteratively until the global training process is successfully completed. This preparatory phase is crucial to computing the BF of the different service providers and establishing how they should act during the service provisioning phase. After this phase, the aggregator saves the computed BF in a shared ledger. In this way, the model should be downloaded and exploited during all subsequent phases.
The forementioned basic ledger can be implemented using any existing lightweight blockchain. The selection of the most appropriate blockchain solution strongly depends on the specific deployment, and, therefore, we consider this choice as orthogonal to our investigation.

\begin{figure*}[ht]
    \centering
    \includegraphics[scale=0.7]{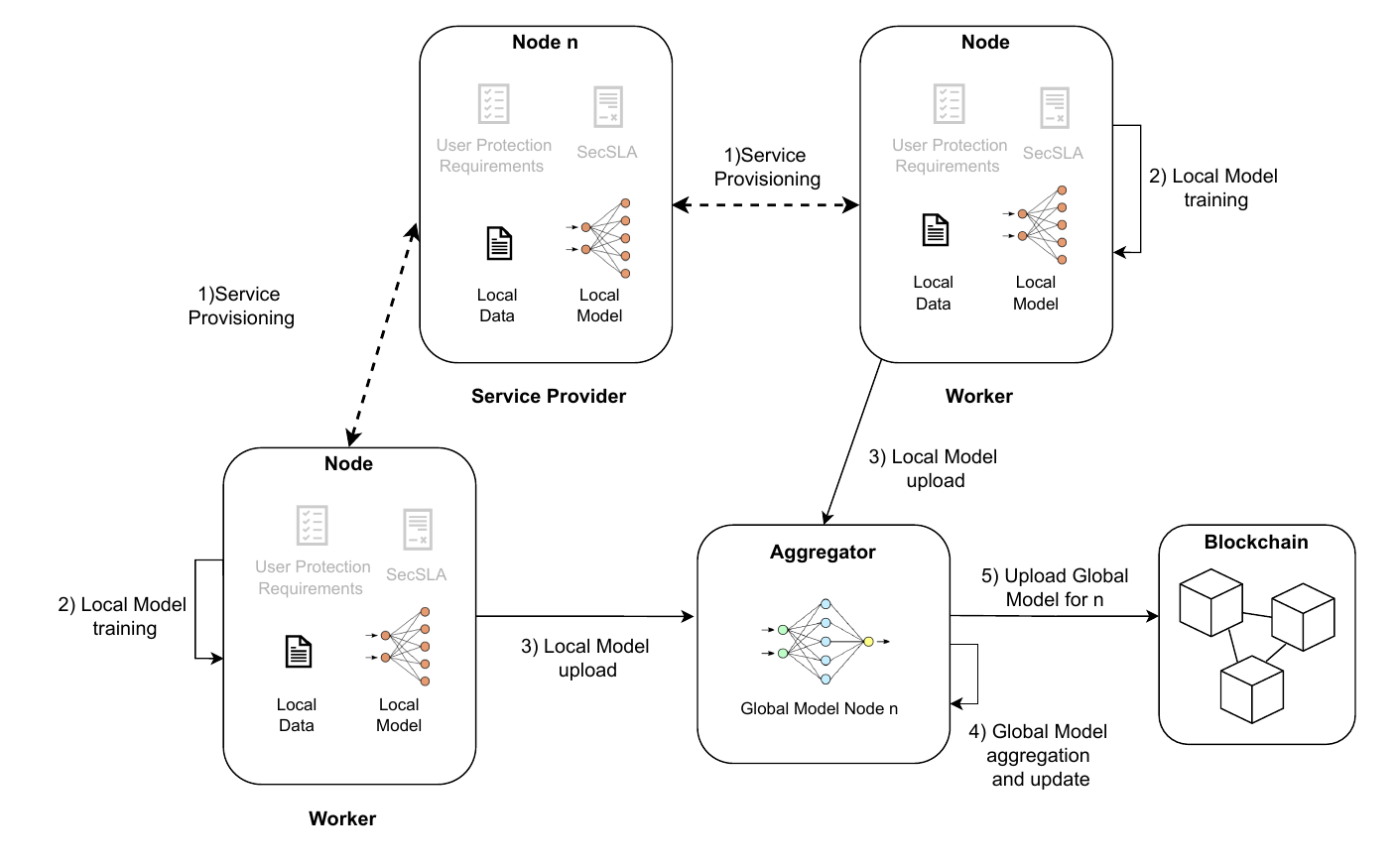}
    \caption{Behavioral Fingerprint Training phase}
    \label{fig:ServiceMonitoringTraining}
\end{figure*}

Figure \ref{fig:ServiceMonitoringTraining} sketches the different actors and interactions of the BF Training phase of our approach.

\begin{figure*}[ht]
    \centering
    \includegraphics[scale=0.7]{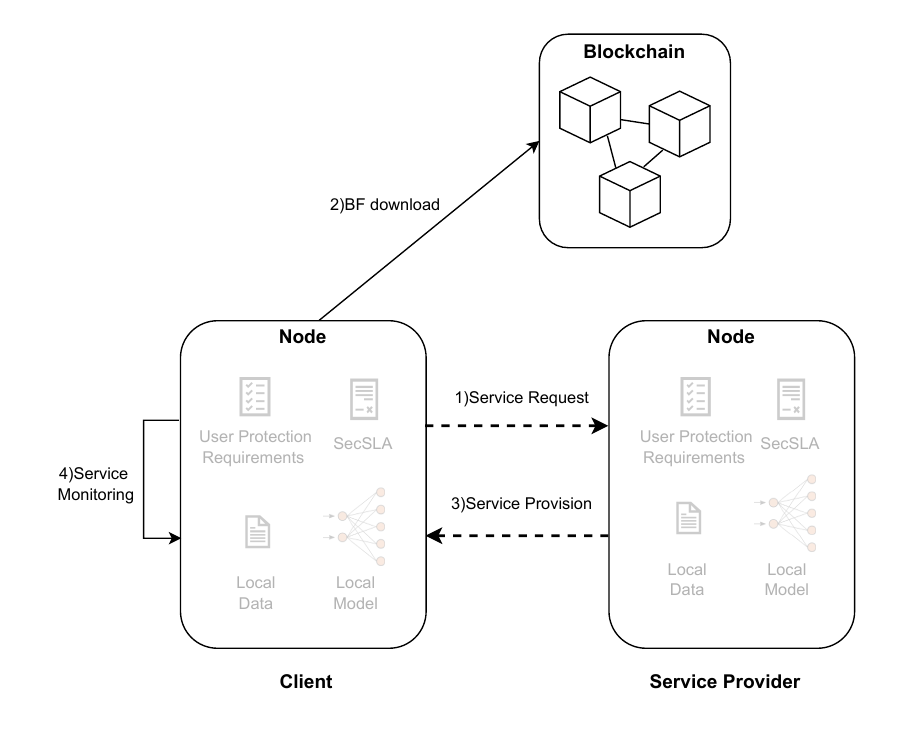}
    \caption{Behavioral Fingerprint Inference phase}
    \label{fig:ServiceMonitoringInference}
\end{figure*}

After the client node performs the selection of the best service provider, the monitoring phase of our framework can start. During this phase, also known as the {\em inference} phase, the learned models are downloaded from the blockchain and used by the clients to infer possible anomalies in the monitored service providers. Figure \ref{fig:ServiceMonitoringInference} illustrates the different actors and interactions of the BF inference phase of our approach.

During each interaction with a service provider $s$, the client compares the observed behavior with the expected one derived from the Behavioral Fingerprinting (BF) model. In particular, a distance $d_s(t)$ is computed between the expected behavioral profile and the observed behavior at time $t$. This distance is then normalized with respect to the maximum possible deviation, obtaining a value $\tilde{d}_s(t) \in [0,1]$.

To ensure robustness over time, these normalized deviations are periodically stored on the blockchain and aggregated by computing their average over a given time window. The trustworthiness score $T_s \in [0,1]$ of a service provider is then defined as:

$$
T_s = 1 - \frac{1}{N} \sum_{t=1}^{N} \tilde{d}_s(t)
$$

\noindent
where $N$ is the number of observations in the considered time window. This formulation ensures that service providers whose behavior closely matches the expected profile achieve higher trustworthiness scores, while deviations from expected behavior lead to lower values of $T_s$.

\section{Experiments}
\label{sec:experiments}
This section describes the experiments carried out to evaluate the effectiveness, robustness, and scalability of the proposed framework. In particular, we first describe the dataset and the simulated IoT marketplace environment used throughout the evaluation, along with the learning and monitoring components of the system. We then introduce the evaluation methodology, including the baselines and evaluation metrics considered. Finally, we present the results obtained together with a critical discussion of the main findings.

\subsection{Dataset}
\label{sub:dataset}

To properly evaluate our solution, we need a scenario in which IoT traffic is available and in which IoT nodes, acting as service providers, violate their SecSLA and change their behavior.
With this in mind, we identified the UNSW IoT Traces dataset~\cite{19sosrIoT}, a publicly available collection of real network traffic from consumer IoT devices in a controlled smart home environment. The dataset contains raw packet captures from 10 IoT devices over two collection periods:
\begin{itemize}
    \item \textbf{Period 1} (May--June 2018): WEMO Motion Sensor, WEMO Power Switch, Samsung Camera, TP-Link Plug, and Netatmo Camera.
    \item \textbf{Period 2} (September--October 2018): Philips Hue Bulb, Amazon Echo, Chromecast, iHome, and LIFX Bulb.
\end{itemize}

Each period consists of benign days, during which devices operate normally, and attack days, during which devices are subjected to various network-layer attacks (ARP Spoofing, TCP SYN Flood, UDP Flood, Ping of Death, SSDP/SNMP Amplification) at different rates (1, 10, and 100 packets/second).
The rationale behind our approach is to use the traffic recorder during the attack days to obtain real communication sequences in which IoT nodes suddenly alter their behavior and therefore no longer conform to the expected operational profile.
Moreover, following the approach outlined in~\cite{aramini2022enhanced}, we pre-process the raw network traffic into per-communication-pair sequences of symbols. Each packet is mapped to a symbolic representation incorporating payload-based features, inter-arrival time, TCP flags, encapsulated protocol type, and packet length. These symbol sequences form the input for training and evaluating Behavioral Fingerprinting (BF) models. In our experiments, we focus on Period~1, using benign days (May~28--31) for BF model training and attack days (June~1--6) to evaluate detection capability and the adaptive response of the system.

\subsection{Experimental Environment}
\label{sub:setup}

We simulate a realistic IoT service marketplace in which multiple service providers are simultaneously available to client agents. Unlike previous approaches that present a single provider per timestep~\cite{arazzi2024deep}, our simulation models the competitive nature of real-world IoT environments where clients must actively select among competing providers.

The marketplace consists of $n$ service providers, each offering one of three service types: temperature monitoring, humidity detection, or air quality measurement. Each provider is characterized by a SecSLA represented as a security class from the CIA triad, a trustworthiness score $T_s \in [0,1]$ (initially set to $1.0$), and an \emph{appearance probability} $p_{\text{app}} \in [0,1]$ that determines its likelihood of being available at any given timestep. At each step, a random subset of providers appears in the marketplace according to their individual appearance probabilities, producing on average $\sum_i p_{\text{app},i} \approx 4$--$5$ available providers per step.

In our default configuration ($n = 10$), the providers are distributed as follows: four temperature providers, three humidity providers, and three air quality providers. The User Protection Requirement is set to $c_{UPR} = [\text{HC}, \text{HI}, \text{HA}]$, representing the highest security demands. The quality of the provider varies significantly within each type of service, with SecSLA vectors ranging from $[3,3,3]$ (premium) to $[1,1,1]$ (low quality). Premium providers have lower appearance probabilities ($p_{\text{app}} = 0.2$--$0.25$), while lower-quality providers appear more frequently ($p_{\text{app}} = 0.45$--$0.70$).

Each agent is assigned an ordered sequence of three tasks to complete: temperature monitoring, humidity detection, and air quality measurement, which must be completed in strict sequential order. At each time step, the agent observes the set of currently available providers and must choose one of $n + 1$ actions: select one of the $n$ providers, or \emph{wait} (i.e., skip the current timestep in anticipation of better options in subsequent steps). An episode ends when all tasks are completed or after a maximum of $T = 20$ steps.

If the agent selects a provider whose service matches the current task and the provider is not compromised, the task is completed, and the agent advances to the next task. If the selected provider is compromised, the task fails silently, the agent receives a negative reward, but the task remains pending, and the SecSLA of the compromised provider remains unchanged. If the selected provider's service does not match the current task, the step is wasted. This design creates a fundamental tension between urgency (limited steps) and caution (avoiding unreliable providers), which the agent must learn to navigate.

The simulation spans $D = 15$ days. Each day, $N_a = 100$ independent agents are deployed in the marketplace. On day $K = 6$ (the \emph{attack day}), a subset of providers becomes compromised. In our default configuration, approximately $20\%$ of the providers are compromised, distributed across different types of services. Compromised providers are assigned higher appearance probabilities ($p_{\text{app}} + 0.2$), simulating the realistic scenario in which adversaries aggressively advertise their services to attract more clients. Crucially, compromised providers continue to advertise valid SecSLAs, making them indistinguishable from benign providers based only on security properties.

\subsection{Learning and Monitoring Components}

This section describes the core components of the proposed framework, including the instantiated Reward Function (based on the definition in Section \ref{sub:ServiceProvisioning}), the DRL agent responsible for service selection, and the BF module used to monitor provider behavior.

The \textbf{Reward Function} is defined as follows. When the agent selects a provider $s$ whose service matches the current task and the provider is not compromised:
$$\mathcal{R}^s_{\text{match}}(t) = 1.0 + \alpha \left(1 - \frac{\delta^s}{\Delta}\right) + (1 - \alpha) \cdot T_s$$
\noindent
where the constant $1.0$ represents the task completion bonus, $\delta^s$ is the Euclidean distance between the provider's SecSLA and the user's requirements, $\Delta$ is the maximum possible distance, $T_s$ is the provider's trustworthiness score, and $\alpha = 0.5$ balances security compliance and provider reliability. When the agent selects a compromised provider:
$$\mathcal{R}_{\text{comp}}(t) = -1.0$$
\noindent
When the agent selects a provider whose service does not match the current task:
$$\mathcal{R}_{\text{mismatch}}(t) = -0.3$$
\noindent
When the agent chooses to wait:
$$\mathcal{R}_{\text{wait}}(t) = -0.05$$
\noindent
Additionally, after successful completion of all tasks, a speed bonus of $3.0 \times (T - t) / T$ is awarded, which incentives efficient completion of tasks. If the episode ends due to the step limit with pending tasks, a penalty of $-1.0 \times |\text{pending}| / |\text{tasks}|$ is applied.

The \textbf{DRL Agent} is implemented using a Deep Q-Network (DQN)~\cite{mnih2015human}. The Q-network consists of two fully connected hidden layers with 128 and 64 neurons, respectively, using ReLU activations, mapping from the observation space to $n + 1$ Q-values (one per provider plus the wait action). The observation is a fixed-size vector of dimensionality $7n + 9$, structured as follows. For each of the $n$ providers, the observation encodes: (i)~an availability flag indicating whether the provider is present in the current market, (ii)~a service match flag indicating whether the provider's service corresponds to the current task, (iii)~the normalized SecSLA vector (3 values), (iv)~the current trustworthiness score $T_s$ and (v)~the precomputed security reward $1 - \delta^s / \Delta$. The global context includes a one-hot encoding of the current task type (3 values), a binary vector of pending tasks (3 values), a normalized completion fraction, the normalized timestep, and a time pressure indicator computed as remaining tasks divided by remaining steps. During both exploration and exploitation, the agent is restricted to valid actions: only available providers and the wait action are considered. Invalid actions (selecting unavailable providers) are masked by setting their Q-values to $-\infty$ before the argmax operation. During $\epsilon$-greedy exploration, random actions are sampled uniformly from the valid set. The agent is pre-trained for $5 \times 10^4$ timesteps on a benign environment (all providers have $T_s = 1.0$) using the Adam optimizer with a learning rate of $10^{-3}$ and discount factor $\gamma = 0.99$. The replay buffer stores up to $10^4$ transitions, and training begins after $200$ initial exploration steps with a batch size of 64. The exploration rate $\epsilon$ decays linearly from $1.0$ to $0.02$ over the first $5 \times 10^3$ steps. The target network is updated every 300 steps. During simulation, the agent continues to learn online: after each simulated day, the DRL policy is updated with $5 \times 10^3$ additional timesteps on the current environment, enabling it to adapt to changes in provider reliability scores.

The \textbf{Behavioral Fingerprinting} model is a GRU-based neural network trained to predict the next symbol in a behavioral sequence, as described in Section~\ref{sub:ServiceMonitoring}. The architecture consists of two GRU layers with 512 and 256 neurons, respectively, followed by a fully connected layer of 128 neurons and a classification output layer. Training is conducted offline during a preparatory phase using benign traffic data (May~28--31) from the UNSW IoT Traces dataset. Following the approach in~\cite{aramini2022enhanced}, raw network packets are assigned to symbolic representations incorporating payload-based features, inter-arrival time, TCP flags, encapsulated protocol types, and packet length. The model is trained using Adam optimizer with a learning rate of $10^{-3}$ and a batch size of $1,000$. During the monitoring phase, the trained BF model is frozen and used in inference mode to compute anomaly scores $\tilde{d}_s(t) \in [0, 1]$ via a sliding window of $w = 250$ symbols. The trustworthiness score is then updated as:
$$T_s = 1 - \frac{1}{N} \sum_{t=1}^{N} \tilde{d}_s(t)$$
\noindent
where $N$ is the number of observations accumulated for providers $s$. This score is shared across agents and stored alongside the provider's profile, enabling collective learning from distributed interactions.

\subsection{Evaluation Methodology}
This section presents the evaluation methodology adopted in our experiments. We first describe the baseline approaches used for comparison, then introduce the metrics employed to assess both service provisioning effectiveness.

To isolate the contribution of each component, we evaluate a baseline with six configurations:

\begin{itemize}
    \item \textbf{B1 -- Random}: The agent selects a provider uniformly at random from the available set at each timestep. No intelligence or learning.
    \item \textbf{B2 -- Greedy (SecSLA only)}: The agent selects the provider available with the highest security reward ($1 - \delta^s / \Delta$) that matches the current task. If no matching provider is available, the agent waits. No learning, no reliability awareness.
    \item \textbf{B3 -- Security-aware}: Same selection criterion as B2 but using the full security reward formulation with $\alpha = 1.0$. This represents the approach of Arazzi et al.~\cite{arazzi2024deep} extended to our marketplace setting, where the agent is aware of security properties but has no mechanism to detect runtime compromise.
    \item \textbf{B4 -- DRL + Local reliability}: A DQN agent with $\alpha = 0.5$ and BF-based monitoring. Each agent maintains its own local trust observations that are merged into a global score at the end of each episode. Reliability information propagates slowly as it depends on individual agent experiences.
    \item \textbf{B5 -- DRL + Shared threshold}: A DQN agent with $\alpha = 0.5$ and shared reliability, but anomaly detection uses a simple threshold in packet-level metrics (detection rate $\approx 30\%$) instead of the BF model. This baseline isolates the contribution of the BF-based detector.
    \item \textbf{B6 -- Full system (ours)}: Complete framework with DQN-based selection, BF-based anomaly detection through Federated Learning, and shared reliability scores between all agents.
\end{itemize}
\begin{figure*}[ht]
\centering
\includegraphics[width=\textwidth]{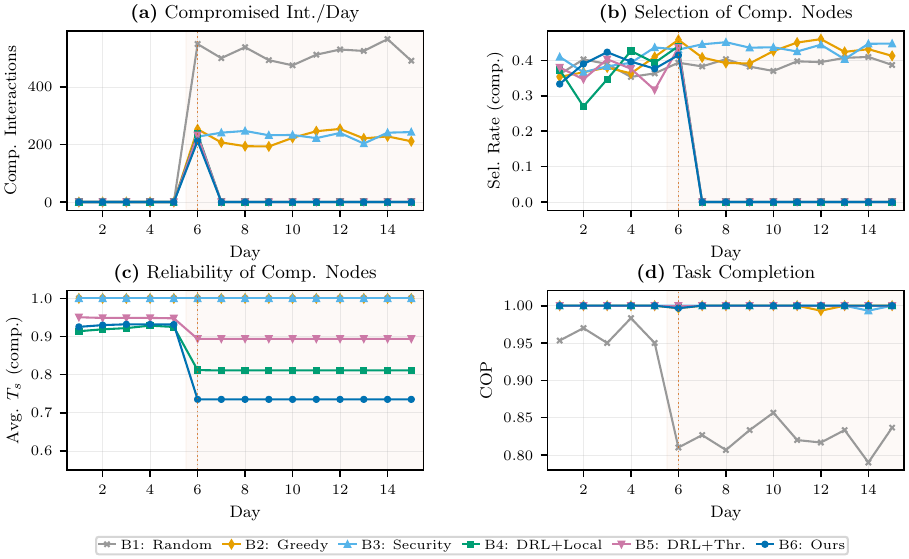}
\caption{DQN marketplace results over the 15-day simulation. The dashed vertical line marks the attack onset at day~6. \textbf{(a)}~Compromised interactions per day: B1--B3 maintain constant exposure ($100$--$300+$ interactions/day), while B4--B6 drop to zero after a single day of adaptation. \textbf{(b)}~Selection rate of compromised providers: B4--B6 reduce selection to $0\%$ by day~7. \textbf{(c)}~Reliability scores of compromised nodes over time. \textbf{(d)}~Wait count: B4--B6 increase waiting behavior as they learn to avoid compromised providers.}
\label{fig:dqn_dashboard}
\end{figure*}

We evaluate each configuration using the following metrics, computed per simulated day and averaged over all agents:

\begin{itemize}
    \item \textbf{COP} (Completed Operations Percentage): the fraction of tasks successfully completed by the agent, averaged over all agents on a given day.
    \item \textbf{Failure Rate}: the fraction of episodes in which at least one task remains uncompleted at episode termination.
    \item \textbf{Compromised Interactions}: the total number of interactions with compromised providers per day, capturing the security risk exposure.
    \item \textbf{Selection Rate}: the proportion of provider selections directed at compromised nodes, measuring how effectively the system avoids them.
    \item \textbf{Reliability Score}: the trustworthiness score $T_s$ of compromised providers over time, reflecting the effectiveness of the monitoring mechanism.
    \item \textbf{Adaptation Speed}: the number of days after the attack onset at which the selection rate of compromised nodes drops below $5\%$.
    \item \textbf{Wait Count}: the number of times agents choose to wait rather than select an available provider, capturing the cost of cautious behavior.
\end{itemize}

\subsection{Results}
\label{sub:results}

We structure our analysis around four central questions: {\em(i)}~Does the DRL agent learn effective service selection in the marketplace? {\em(ii)}~Does the reliability signal emerge and enable adaptation after compromise? {\em(iii)}~How do the different monitoring strategies compare? {\em(iv)}~How does the framework scale with the size of the network?

All experiments are conducted in the marketplace environment described in Section~\ref{sub:setup}. Unless otherwise stated, results are reported for the default configuration ($n = 10$ providers, $N_a = 100$ agents per day, $D = 15$ days, attack onset at day $K = 6$).

\subsubsection{Pre-Attack Performance}

During the benign period (days 1--5), all DRL-based baselines (B4--B6) achieve COP~$= 1.0$ with a zero failure rate, demonstrating that the DQN agent successfully learns to navigate the market and completes all tasks in sequential order. The agent completes its three tasks in approximately 4--5 steps on average, using the remaining steps as waits when no suitable provider is available. The pre-training phase ($5 \times 10^4$ timesteps) is sufficient for the agent to learn effective provider selection with action masking.

Baselines B2 (Greedy) and B3 (Security-aware) also achieve COP~$= 1.0$ during this period, since they deterministically select the best available provider that matches the current task. B1 (Random) achieves a lower COP of approximately $0.93$--$0.97$ during benign days due to its uninformed selection strategy, which frequently picks providers whose services do not match the current task or chooses to interact with low-quality providers, wasting steps.

\subsubsection{Attack Response and Adaptation}

\begin{figure*}[ht]
\centering
\includegraphics[width=\textwidth]{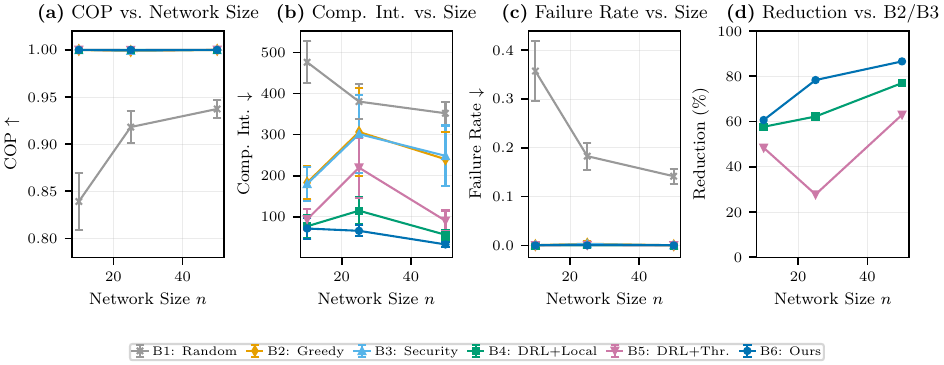}
\caption{Scalability analysis across network sizes $n \in \{10, 25, 50\}$ (mean $\pm$ std over 10 seeds). \textbf{(a)}~COP vs.\ network size: B1 degrades while B4--B6 maintain $1.0$. \textbf{(b)}~Compromised interactions vs.\ network size: B6 reduces interactions by $61\%$--$87\%$ relative to B2/B3. \textbf{(c)}~Failure rate vs.\ network size. \textbf{(d)}~Relative reduction in compromised interactions for monitoring-enabled baselines compared to B2/B3.}
\label{fig:ablation}
\end{figure*}
On day 6, the compromised providers become active. The immediate effect is visible on all baselines: agents that select compromised providers receive a reward of $-1.0$ and do not complete the associated task. The critical difference lies in what happens \emph{after} this initial exposure.

\paragraph{B1 -- Random} The random agent continues to select compromised providers at a constant rate throughout the post-attack period. With approximately $20\%$ of providers compromised and higher appearance probabilities for adversarial nodes, B1 accumulates $400$--$550$ compromised interactions per day. Its COP degrades to approximately $0.80$--$0.85$ as failed interactions consume steps that could otherwise be used for the completion of legitimate tasks, resulting in a failure rate of $44\%$.

\paragraph{B2 and B3 -- Greedy / Security-aware} These baselines select providers based solely on SecSLA quality. Since compromised providers retain valid SecSLAs, B2 and B3 continue to select them whenever they offer the best security match for the current task. Both baselines accumulate $200$--$250$ compromised interactions per day with no adaptation over time. Their COP remains at $0.999$ because alternative providers are available in subsequent steps, but the security exposure is severe and persistent.

\paragraph{B4 -- DRL + Local reliability} The DQN agent with local BF-based monitoring demonstrates rapid adaptation. On day~6, the agent encounters compromised providers and receives the $-1.0$ penalty. The BF monitor detects behavioral anomalies and begins updating the local trustworthiness scores. After the online training phase ($5 \times 10^3$ timesteps at the end of day~6), the updated Q-values reflect the negative experience. By day~7, the selection rate of all compromised providers drops to \textbf{zero}, and compromised interactions cease entirely. The reliability scores of compromised nodes drop to $T_s \approx 0.80$--$0.83$, reflecting the local monitoring observations. The agent compensates by increasing its wait count from $\sim\!45$ (pre-attack) to $\sim\!190$--$230$ (post-attack), choosing to skip timesteps when only compromised providers are available for the current task.

\paragraph{B5 -- DRL + Shared threshold} The behavior of B5 closely mirrors B4. Despite using a simpler threshold-based detector (detection rate $\approx 30\%$), the DQN agent learns to avoid compromised providers equally quickly, within a single day. The key insight is that the DQN's adaptation is driven primarily by the \emph{direct reward signal} ($-1.0$ for compromised interactions) rather than the reliability score. The threshold detector produces weaker reliability updates ($T_s \approx 0.88$--$0.92$), but the online Q-learning is sufficient to encode avoidance behavior. Compromised interactions drop to zero from day~7 onward.

\paragraph{B6 -- Full system} Our complete framework achieves the same rapid adaptation as B4 and B5, with the additional benefit of the strongest reliability signal. On day~6, B6 accumulates 212 compromised interactions as the DQN encounters adversarial providers for the first time. The BF-based detector with shared observations immediately updates the trustworthiness scores, producing the deepest reliability drop among all monitoring baselines: $T_{N1} = 0.77$, $T_{N5} = 0.78$, and $T_{N8} = 0.65$ after a single day of exposure, compared to $T_s \approx 0.80$--$0.83$ for B4 and $T_s \approx 0.88$--$0.92$ for B5. By day~7, the selection rate of all compromised providers drops to zero, and the agent maintains COP~$= 1.0$ for the remainder of the simulation. The wait count increases from $\sim\!45$ (pre-attack) to $\sim\!190$ (post-attack), reflecting the agent's learned strategy of waiting for benign alternatives.

\subsubsection{Quantitative Comparison}

Table~\ref{tab:dqn_results} summarizes the post-attack metrics for all baselines.

\begin{table*}[ht]
\centering
\caption{Post-attack average metrics ($d \geq 6$) in the marketplace environment with DQN agents. Compromised interactions (Comp.Int.) are averaged per day. Adaptation speed indicates days until compromised selection rate drops below $5\%$. Best values in \textbf{bold}.}
\label{tab:dqn_results}
\begin{tabular}{lccccc}
\toprule
\textbf{Method} & \textbf{COP}$\uparrow$ & \textbf{Fail}$\downarrow$ & \textbf{Comp.Int.}$\downarrow$ & \textbf{Waits} & \textbf{Adapt.} \\
\midrule
B1: Random          & 0.823 & 0.440 & 518   & 322   & never \\
B2: Greedy          & 0.999 & 0.003 & 223   & 101   & never \\
B3: Security        & 0.999 & 0.003 & 233   & 101   & never \\
B4: DRL+Local       & \textbf{1.000} & \textbf{0.000} & 24  & 184  & \textbf{1 day} \\
B5: DRL+Shared Thr. & \textbf{1.000} & \textbf{0.000} & 23  & 185  & \textbf{1 day} \\
B6: Full (Ours)     & \textbf{1.000} & \textbf{0.000} & \textbf{21}  & 182  & \textbf{1 day} \\
\bottomrule
\end{tabular}
\end{table*}

The results reveal several key findings:

\begin{enumerate}
    \item \textbf{DRL enables instant adaptation.} All DRL-based baselines (B4--B6) reduce compromised interactions to zero within a single day of the attack onset. The online learning mechanism ($5 \times 10^3$ timesteps per day) is sufficient for the DQN to update its policy based on the negative reward signal from compromised interactions. In contrast, B1--B3 lack any learning mechanism and continue to expose agents to compromised providers indefinitely.

    \item \textbf{Monitoring quality affects reliability depth, not adaptation speed.} While B4, B5, and B6 all achieve the same adaptation speed (1 day), they differ in the depth of the reliability signal. B6 (BF-based, shared) achieves the lowest reliability scores for compromised providers, followed by B4 (BF-based, local) and B5 (threshold-based, shared). This difference becomes significant in scenarios involving provider re-evaluation or dynamic trust recovery, where a deeper reliability drop provides a stronger safety margin.

    \item \textbf{Waiting is the cost of safety.} The DRL baselines compensate for avoiding compromised providers by increasing their wait count. Pre-attack, agents wait approximately 45--55 times per day; post-attack, this increases to 150--230. This reflects the agent's learned strategy: when the only available provider for the current task is compromised, the agent waits for a benign alternative rather than accepting the risk. Despite the increased waiting, COP remains at $1.0$, demonstrating that the $T = 20$ step budget provides sufficient slack for cautious behavior.

    \item \textbf{Security-blind baselines pose silent risks.} B2 and B3 achieve COP~$\approx 1.0$ but accumulate $220$+ compromised interactions per day. This highlights a critical limitation of SecSLA-only approaches: task completion metrics alone are insufficient to assess the security posture of the system. {\em Without behavioral monitoring, agents continuously interact with compromised providers and therefore expose users to data leakage, manipulation, and other attack consequences that are not captured by COP.}
\end{enumerate}

\subsubsection{Ablation: Component Contributions}

The progression from B3 to B6 isolates the contribution of each framework component:

\begin{itemize}
    \item \textbf{B3 $\to$ B4} (adding DRL + local BF monitoring): The most significant improvement. The DRL agent gains the ability to learn from negative interactions and adapt its policy online. The BF monitor provides local reliability updates that reinforce learned avoidance behavior. Compromised interactions drop from $228$/day to $24$/day (an $89\%$ reduction).

    \item \textbf{B4 $\to$ B5} (sharing reliability, simpler detector): Sharing trust observations across agents does not significantly accelerate adaptation in this setting, as the DQN's online learning already converges within a single day based on direct reward signals. The threshold detector produces weaker reliability scores, but achieves comparable avoidance behavior, suggesting that the DRL's reward-driven learning is the primary adaptation mechanism.

    \item \textbf{B5 $\to$ B6} (BF-based detection with sharing): Replacing the threshold detector with the GRU-based BF model produces the deepest reliability drop. Although adaptation speed remains the same (1 day), the BF model's ability to detect subtle behavioral anomalies results in more accurate trustworthiness scores, providing stronger long-term protection and more informative signals for system-level decision-making.
\end{itemize}

\subsubsection{Scalability Analysis}
To evaluate the robustness of our framework on different network scales, we conduct an ablation study varying the number of providers $n \in \{10, 25, 50\}$. For each network size, the number of providers per service type and the number of compromised nodes are scaled proportionally ($\sim\!20\%$ compromised). Each configuration is evaluated over 10 independent random seeds, and we report the mean and standard deviation of all metrics. This study assesses whether the advantages of BF-based monitoring and shared reliability persist, and potentially amplify, as the network grows and the provider landscape becomes more complex.
Table~\ref{tab:ablation_size} presents the results of the scalability study, where we vary the network size $n \in \{10, 25, 50\}$ and report metrics averaged over 10 independent seeds. These experiments use the greedy agent variant to enable rapid evaluation across the parameter space.

\begin{table}[ht]
\centering
\scriptsize
\caption{Scalability study: post-attack metrics across network sizes (mean $\pm$ std over 10 seeds). Comp.Int.\ denotes average daily compromised interactions.}
\label{tab:ablation_size}
\begin{tabular}{clccc}
\toprule
\textbf{$n$} & \textbf{Method} & \textbf{COP} $\uparrow$ & \textbf{Comp.Int.} $\downarrow$ & \textbf{Fail} $\downarrow$ \\
\midrule
10 & B1: Random          & $0.839 \pm 0.031$ & $476 \pm 51$  & $0.357 \pm 0.061$ \\
   & B2: Greedy          & $1.000 \pm 0.000$ & $183 \pm 40$  & $0.000 \pm 0.000$ \\
   & B3: Security        & $1.000 \pm 0.000$ & $180 \pm 41$  & $0.000 \pm 0.001$ \\
   & B4: Local Rel.      & $1.000 \pm 0.000$ & $77 \pm 27$   & $0.000 \pm 0.000$ \\
   & B5: Shared Thr.     & $1.000 \pm 0.000$ & $94 \pm 25$   & $0.000 \pm 0.000$ \\
   & \textbf{B6: Full}   & $1.000 \pm 0.000$ & $\mathbf{71 \pm 26}$   & $0.000 \pm 0.000$ \\
\midrule
25 & B1: Random          & $0.918 \pm 0.017$ & $380 \pm 42$  & $0.183 \pm 0.028$ \\
   & B2: Greedy          & $0.999 \pm 0.003$ & $306 \pm 106$ & $0.003 \pm 0.007$ \\
   & B3: Security        & $0.999 \pm 0.002$ & $301 \pm 96$  & $0.003 \pm 0.005$ \\
   & B4: Local Rel.      & $1.000 \pm 0.000$ & $115 \pm 34$  & $0.000 \pm 0.000$ \\
   & B5: Shared Thr.     & $1.000 \pm 0.000$ & $220 \pm 73$  & $0.001 \pm 0.001$ \\
   & \textbf{B6: Full}   & $1.000 \pm 0.000$ & $\mathbf{66 \pm 14}$   & $0.000 \pm 0.000$ \\
\midrule
50 & B1: Random          & $0.937 \pm 0.009$ & $352 \pm 28$  & $0.142 \pm 0.015$ \\
   & B2: Greedy          & $1.000 \pm 0.000$ & $240 \pm 66$  & $0.000 \pm 0.000$ \\
   & B3: Security        & $1.000 \pm 0.000$ & $248 \pm 74$  & $0.000 \pm 0.001$ \\
   & B4: Local Rel.      & $1.000 \pm 0.000$ & $56 \pm 12$   & $0.000 \pm 0.000$ \\
   & B5: Shared Thr.     & $1.000 \pm 0.000$ & $90 \pm 25$   & $0.000 \pm 0.000$ \\
   & \textbf{B6: Full}   & $1.000 \pm 0.000$ & $\mathbf{33 \pm 6}$    & $0.000 \pm 0.000$ \\
\bottomrule
\end{tabular}
\end{table}

The scalability results reveal three important trends:

\begin{enumerate}
    \item \textbf{B6's advantage amplifies with scale.} As the network grows from $n = 10$ to $n = 50$, B6 reduces its compromised interactions from $71$ to $33$ per day, a reduction of $61\%$ at $n = 10$, $78\%$ at $n = 25$, and $87\%$ at $n = 50$ relative to B2/B3. Larger networks provide more benign alternatives, which B6 effectively leverages through its reliability-aware selection.

    \item \textbf{B1's performance improves but remains inferior.} The random baseline benefits from larger networks (more providers increase the probability of randomly selecting a benign one), with COP improving from $0.839$ to $0.937$. However, it still accumulates $352$ compromised interactions per day at $n = 50$, demonstrating that scale alone does not solve the security problem without intelligent selection.

    \item \textbf{Monitoring consistency.} B4 and B6 maintain COP~$= 1.0$ with zero failure rate across all network sizes, confirming that the monitoring and reliability mechanisms generalize effectively. The standard deviations remain small ($\pm 6$--$\pm 27$ for compromised interactions), indicating consistent performance across random seeds.
\end{enumerate}

\subsubsection{Discussion}

Our experimental evaluation demonstrates that the combination of DRL-based service selection and BF-based behavioral monitoring provides an effective defense against compromised service providers in IoT marketplaces. The DQN agent's ability to learn online from negative interactions enables rapid adaptation, within a single day of attack onset, while maintaining perfect task completion rates. The BF monitoring component adds depth to the reliability signal, producing trustworthiness scores that accurately reflect provider behavior and support system-level security decisions beyond immediate avoidance.

A notable finding is the role of the design of the marketplace environment in enabling a meaningful evaluation. Unlike single-provider-per-step formulations, the marketplace setting creates genuine decision-making challenges: agents must balance security, quality, and urgency when selecting among competing providers, and the option to wait introduces a meaningful trade-off between safety and efficiency. This more realistic formulation ensures that the experimental results reflect the complexities of real-world IoT service provisioning.

The scalability analysis further confirms that our framework's advantages are not limited to small networks. As the provider landscape becomes more complex, the relative benefit of reliability-aware selection increases, suggesting that the approach is well-suited for large-scale IoT deployments such as smart cities and industrial IoT environments.

\section{Conclusion}
\label{sec:conclusion}
As the Internet of Things (IoT) ecosystem continues to expand, ensuring secure service provisioning is critical for maintaining the reliability, functionality, and security of interconnected devices and services that are increasingly exposed to sophisticated cyber threats.

In this work, we introduced a comprehensive framework for secure service provisioning that jointly addresses smart object selection and behavioral monitoring. By integrating Deep Reinforcement Learning (DRL) for adaptive decision-making and Federated Learning (FL) for distributed Behavioral Fingerprinting (BF), our approach enables the selection of service providers not only based on compliance with user-defined security requirements but also on their observed trustworthiness over time. In particular, the proposed mechanism incorporates a trust score derived from BF into the decision process, allowing the system to dynamically balance functional suitability, security constraints, and reliability.

The adoption of FL ensures that behavioral models can be collaboratively learned across distributed nodes while preserving data privacy, while DRL allows the system to operate effectively in dynamic and uncertain environments, continuously improving its selection policy through interaction. Experimental results demonstrate that our framework achieves robust performance, effectively identifying reliable service providers while maintaining scalability and low computational overhead. In particular, the solution proves to be suitable even for resource-constrained IoT devices, highlighting its practical applicability.

These findings highlight the potential of our framework to strengthen IoT security by mitigating risks associated with service provisioning while preserving system performance. Future work will focus on extending the framework to more complex and large-scale IoT scenarios, incorporating additional contextual factors (e.g., energy consumption, latency, and trust dynamics), and investigating more advanced learning strategies to further enhance adaptability and resilience against evolving threats.


{\footnotesize
\begin{thebibliography}{10}
\expandafter\ifx\csname url\endcsname\relax
  \def\url#1{\texttt{#1}}\fi
\expandafter\ifx\csname urlprefix\endcsname\relax\def\urlprefix{URL }\fi
\expandafter\ifx\csname href\endcsname\relax
  \def\href#1#2{#2} \def\path#1{#1}\fi

\bibitem{adat2018security}
V.~Adat, B.~B. Gupta, Security in internet of things: issues, challenges,
  taxonomy, and architecture, Telecommunication Systems 67~(3) (2018) 423--441.

\bibitem{arazzi2025privacy}
M.~Arazzi, M.~Cihangiroglu, S.~Nicolazzo, A.~Nocera, A privacy-preserving and
  biometric-aware tasks reallocation strategy in industry 5.0, in: 2025 IEEE
  30th International Conference on Emerging Technologies and Factory Automation
  (ETFA), IEEE, 2025, pp. 1--8.

\bibitem{alghamdi2019secure}
T.~A. Alghamdi, I.~Ali, N.~Javaid, M.~Shafiq, Secure service provisioning
  scheme for lightweight iot devices with a fair payment system and an
  incentive mechanism based on blockchain, IEEE Access 8 (2019) 1048--1061.

\bibitem{rios2022security}
E.~Rios, M.~Higuero, X.~Larrucea, M.~Rak, V.~Casola, E.~Iturbe, Security and
  privacy service level agreement composition for internet of things systems on
  top of standard controls, Computers \& Electrical Engineering 98 (2022)
  107690.

\bibitem{casola2018security}
V.~Casola, A.~De~Benedictis, M.~Rak, U.~Villano, A security metric catalogue
  for cloud applications, in: Complex, Intelligent, and Software Intensive
  Systems: Proceedings of the 11th International Conference on Complex,
  Intelligent, and Software Intensive Systems (CISIS-2017), Springer, 2018, pp.
  854--863.

\bibitem{nicolazzo2026service}
S.~Nicolazzo, A.~Nocera, W.~Pedrycz, Service level agreement (sla) and security
  sla (secsla): A comprehensive survey, Journal of Network and Systems
  Management 34~(3) (2026) 74.

\bibitem{arazzi2024deep}
M.~Arazzi, S.~Nicolazzo, A.~Nocera, A deep reinforcement learning approach for
  security-aware service acquisition in iot, Journal of Information Security
  and Applications (2024).

\bibitem{aramini2022enhanced}
A.~Aramini, M.~Arazzi, T.~Facchinetti, L.~S. Ngankem, A.~Nocera, An enhanced
  behavioral fingerprinting approach for the internet of things, in: 2022 IEEE
  18th International Conference on Factory Communication Systems (WFCS), IEEE,
  2022, pp. 1--8.

\bibitem{ferretti2021h2o}
M.~Ferretti, S.~Nicolazzo, A.~Nocera, {H2O: Secure Interactions in IoT via
  Behavioral Fingerprinting}, Future Internet 13~(5) (2021) 117.

\bibitem{arazzi2023fully}
M.~Arazzi, S.~Nicolazzo, A.~Nocera, A fully privacy-preserving solution for
  anomaly detection in iot using federated learning and homomorphic encryption,
  Information Systems Frontiers (2023) 1--24.

\bibitem{deng2018composition}
S.~Deng, Z.~Xiang, J.~Yin, J.~Taheri, A.~Y. Zomaya, Composition-driven iot
  service provisioning in distributed edges, IEEE Access 6 (2018) 54258--54269.

\bibitem{niemirepo2015service}
T.~Niemirepo, M.~Sihvonen, V.~Jordan, J.~Heinil{\"a}, Service platform for
  automated iot service provisioning, in: 2015 9th International Conference on
  Innovative Mobile and Internet Services in Ubiquitous Computing, IEEE, 2015,
  pp. 325--329.

\bibitem{zhao2016event}
S.~Zhao, L.~Yu, B.~Cheng, An event-driven service provisioning mechanism for
  iot (internet of things) system interaction, IEEE Access 4 (2016) 5038--5051.

\bibitem{khan2017towards}
Z.~Khan, Z.~Pervez, A.~G. Abbasi, Towards a secure service provisioning
  framework in a smart city environment, Future Generation Computer Systems 77
  (2017) 112--135.

\bibitem{shahidinejad2023blockchain}
A.~Shahidinejad, J.~Abawajy, Blockchain-based self-certified key exchange
  protocol for hybrid electric vehicles, IEEE Transactions on Consumer
  Electronics (2023).

\bibitem{kazim2018framework}
M.~Kazim, L.~Liu, S.~Y. Zhu, A framework for orchestrating secure and dynamic
  access of iot services in multi-cloud environments, IEEE Access 6 (2018)
  58619--58633.

\bibitem{henning1999security}
R.~R. Henning, Security service level agreements: quantifiable security for the
  enterprise?, in: Proceedings of the 1999 workshop on New security paradigms,
  ACM, Ontario, Canada, 1999, pp. 54--60.

\bibitem{enisa2009enisa}
S.~ENISA, Enisa (2009).

\bibitem{casola2016automatically}
V.~Casola, A.~De~Benedictis, M.~Era{\c{s}}cu, J.~Modic, M.~Rak, Automatically
  enforcing security slas in the cloud, IEEE Transactions on Services Computing
  10~(5) (2016) 741--755.

\bibitem{chan2004role}
C.~K. Chan, U.~Chandrashekhar, S.~H. Richman, S.~R. Vasireddy, The role of slas
  in reducing vulnerabilities and recovering from disasters, Bell Labs
  Technical Journal 9~(2) (2004) 189--203.

\bibitem{keller2003wsla}
A.~Keller, H.~Ludwig, The wsla framework: Specifying and monitoring service
  level agreements for web services, Journal of Network and Systems Management
  11 (2003) 57--81.

\bibitem{maarouf2015review}
A.~Maarouf, A.~Marzouk, A.~Haqiq, A review of sla specification languages in
  the cloud computing, in: 2015 10th International Conference on Intelligent
  Systems: Theories and Applications (SITA), IEEE, Rabat Morocco, 2015, pp.
  1--6.

\bibitem{ludwig2003web}
H.~Ludwig, A.~Keller, A.~Dan, R.~P. King, R.~Franck, Web service level
  agreement (wsla) language specification, Ibm corporation (2003) 815--824.

\bibitem{bianco2008service}
P.~Bianco, G.~A. Lewis, P.~Merson, Service level agreements in service-oriented
  architecture environments, Carnegie Mellon University, Software Engineering
  Institute, 2008.

\bibitem{zhang2021survey}
C.~Zhang, Y.~Xie, H.~Bai, B.~Yu, W.~Li, Y.~Gao, A survey on federated learning,
  Knowledge-Based Systems 216 (2021) 106775.

\bibitem{arulkumaran2017deep}
K.~Arulkumaran, M.~P. Deisenroth, M.~Brundage, A.~A. Bharath, Deep
  reinforcement learning: A brief survey, IEEE Signal Processing Magazine
  34~(6) (2017) 26--38.

\bibitem{konevcny2015federated}
J.~Kone{\v{c}}n{\`y}, B.~McMahan, D.~Ramage, Federated optimization:
  Distributed optimization beyond the datacenter, arXiv preprint
  arXiv:1511.03575 (2015).

\bibitem{nguyen2021federated}
D.~C. Nguyen, M.~Ding, P.~N. Pathirana, A.~Seneviratne, J.~Li, H.~V. Poor,
  Federated learning for internet of things: A comprehensive survey, IEEE
  Communications Surveys \& Tutorials 23~(3) (2021) 1622--1658.

\bibitem{sanchez2021survey}
P.~M.~S. S{\'a}nchez, J.~M.~J. Valero, A.~H. Celdr{\'a}n, G.~Bovet, M.~G.
  P{\'e}rez, G.~M. P{\'e}rez, A survey on device behavior fingerprinting: Data
  sources, techniques, application scenarios, and datasets, IEEE Communications
  Surveys \& Tutorials 23~(2) (2021) 1048--1077.

\bibitem{19sosrIoT}
A.~Hamza, H.~Habibi~Gharakheili, T.~A. Benson, V.~Sivaraman,
  \href{https://www2.ee.unsw.edu.au/~hhabibi/publications.html#19sosrIoT}{{Detecting
  Volumetric Attacks on LoT Devices via SDN-Based Monitoring of MUD Activity}},
  in: Proc. ACM SOSR, San Jose, CA, USA, 2019.
\newblock \href {https://doi.org/10.1145/3314148.3314352}
  {\path{doi:10.1145/3314148.3314352}}.
\newline\urlprefix\url{https://www2.ee.unsw.edu.au/~hhabibi/publications.html#19sosrIoT}

\bibitem{mnih2015human}
V.~Mnih, K.~Kavukcuoglu, D.~Silver, A.~A. Rusu, J.~Veness, M.~G. Bellemare,
  A.~Graves, M.~Riedmiller, A.~K. Fidjeland, G.~Ostrovski, et~al., Human-level
  control through deep reinforcement learning, nature 518~(7540) (2015)
  529--533.

\end{thebibliography}
}
\end{document}